\documentclass[english,aps, prb, preprint, showpacs, floatfix]{revtex4}
\usepackage[T1]{fontenc}
\usepackage[latin1]{inputenc}
\usepackage{graphicx}
\makeatletter
\usepackage{babel}
\makeatother
\begin{document}

\title{Strain Effect on Energy Gaps of Armchair Graphene Nanoribbons}

\author{Lian Sun}
\affiliation{Hefei National Laboratory for Physical Sciences at
Microscale, University of Science and Technology of China, Hefei,
Anhui 230026, P.R. China}

\author{Qunxiang Li}
\affiliation{Hefei National Laboratory for Physical Sciences at
Microscale, University of Science and Technology of China, Hefei,
Anhui 230026, P.R. China}

\author{Hao Ren}
\affiliation{Hefei National Laboratory for Physical Sciences at
Microscale, University of Science and Technology of China, Hefei,
Anhui 230026, P.R. China}

\author{Q. W. Shi}
\affiliation{Hefei National Laboratory for Physical Sciences at
Microscale, University of Science and Technology of China, Hefei,
Anhui 230026, P.R. China}

\author{Jinlong Yang}\thanks{Corresponding author. E-mail: jlyang@ustc.edu.cn}
\affiliation{Hefei National Laboratory for Physical Sciences at
Microscale, University of Science and Technology of China, Hefei,
Anhui 230026, P.R. China}

\author{J. G. Hou}
\affiliation{Hefei National Laboratory for Physical Sciences at
Microscale, University of Science and Technology of China, Hefei,
Anhui 230026, P.R. China}

\date{\today}

\begin{abstract}
We report a first-principles study on electronic structures of the
deformed armchair graphene nanoribbons (AGNRs). The variation of the
energy gap of AGNRs as a function of uniaxial strain displays a
zigzag pattern, which indicates that the energy gaps of AGNRs can be
effectively tuned. The spatial distributions of two occupied and two
empty subbands close to the Fermi level are swapped under different
strains. The tunable width of energy gaps becomes narrower as
increasing the width of AGNRs. Our simulations with tight binding
approximation, including the nearest neighbor hopping integrals
between $\pi$- orbitals of carbon atoms, reproduce these results by
first-principles calculations. One simple empirical formula is
obtained to describe the scaling behavior of the maximal value of
energy gap as a function of the width of AGNRs.
\end{abstract}


\maketitle

Nanoscale carbon materials including fullerenes and carbon nanotubes
have attracted a great deal of research interest owing to its
versatile electronic properties.\cite{fullerenes,cnt} Among them,
graphene fabricated by Novoselov \emph{et al.} firstly has been
studied extensively.\cite{Novoselov-Nature} Many interesting
properties of this kind layered two-dimensional carbon
nanostructure, such as the Landau quantization,\cite{Matsui} the
integer quantum-Hall effect,\cite{Zhang,Dmitry,Kane,Yuanbozhang} and
the quantization minimum conductivity,\cite{Tworzydlo} have been
investigated by several experimental and theoretical research
groups. Now much attention has focused on graphene nanoribbons
(GNRs) with various widths, which can be realized by cutting the
exfoliated graphene, or by patterning graphene
epitaxially.\cite{Fujita,Yuanbozhang,Berger} The edge carbon atoms
of graphene ribbons have two typical topological shapes: namely
armchair and zigzag. All zigzag graphene nanoribbons (ZGNR) are
metallic due to a localized state at the Fermi level. This has been
confirmed by scanning tunneling spectroscopy and atomic force
microscopy.\cite{Y.Kobayashi,Y.Niimi,S.Banerjee} It originates from
a gauge field produced by lattice deformation. Such a localized
state, however, does not appear in AGNRs. Fujita \emph{et al.} have
calculated the energy band structure for ZGNRs and AGNRs by using
tighting-binding approximations (TBA) for the $\pi$-states of
carbon.\cite{Fujita} They have found that when the width ($W$) of
graphene ribbon is 3n-1, where n is an integer, AGNR is metallic;
otherwise it is semiconducting. However, the first-principles
calculations have shown that the hydrogen passivated AGNRs always
have nonzero and direct band gaps at the local (spin) density
approximation level.\cite{Son} The energy gaps (E$_{g}$) of AGNRs as
a function of ribbon width are classified into three families, in
which E$_{g}$(W=3n+1)$>$E$_{g}$(W=3n)$>$E$_{g}$(W=3n+2).\cite{Son}

The capability to control GNRs' electronic properties are highly
desired to build future nanodevice directly on GNRs. For example,
the electronic structures of AGNRs can be altered through the
chemical edge modification.\cite{Wang-GNR} Another possible
effective way is to apply external strain, since previous studies
have indicated that the uniaxial strain affected significantly the
electronic properties of nanoscale carbon
material.\cite{PRB-Heyd,PRL-Yang,SS-Ito} Existing theoretical works
focuse on electronic structure and magnetic properties of
GNRs,\cite{Peres,Wakabayashi} nevertheless, more attention should be
paid to the geometric deformation effect. In this paper, we perform
\emph{ab inito} calculation about strain effect on the electronic
structure of AGNRs with various widths. Theoretical results show
that the energy gaps can be tuned effectively by external strain.
The tunable width of energy gap decreases when the width of AGNRs
increases. The nearest neighbor hopping integrals between $\pi$-
orbitals of carbon atoms are responsible for the variation of energy
gap under uniaxial strain.

We apply density functional theory with generalized gradient
approximation (GGA) implemented with the DMol$^{3}$
package.\cite{dmol} The Becke exchange gradient correction and the
Lee-Yang-Parr correlation gradient correction are
adopted.\cite{blyp} The basis set consists of the double numerical
atomic orbitals augmented by polarization functions. The
calculations are all-electron ones with scalar relativistic
corrections. Self-consistent field procedure is carried out with a
convergence criterion of 5.0$\times$10$^{-5}$ atomic units (a.u.) on
the energy and electron density. Geometry optimizations are
conducted with convergence criterions of 5$\times$10$^{-3}$ on the
gradient, 5$\times$10$^{-3}$ on the displacement, and
5$\times$10$^{-5}$ a.u. on the energy. Medium grid mesh points are
employed for the matrix integrations, the real-space global cutoff
radius of all atoms is set to be 5.5 \AA, and uniformly 22 K points
along the one dimensional Brillouin zone are used to calculate
electronic structures of the AGNRs.

Here, AGNRs with widths W=12, 13, and 14 are chosen to represent three
typical families (corresponding to 3n, 3n+1, and 3n+2, respectively),
similar to the previous theoretical study.\cite{Son} As an example,
the schematic of an AGNR with width W=13 is shown in Figure~\ref{fig1}.
To avoid the effects of the $\sigma$ electronic states near the Fermi
level, the dangling bonds of edge carbon atom are saturated by one
hydrogen atom. All atomic positions of AGNRs atoms are allowed to
relax by using a rectangular supercell, in which AGNR is set with
its edge separated by at least 10 {\AA} from neighboring AGNRs.
As a benchmark, the electronic structures of AGNRs without geometric
deformation are calculated. The calculated energy gaps are 0.55, 0.90,
and 0.19 eV for the AGNR with width W=12, 13, and 14, respectively,
which reproduce the previous DFT results.\cite{Son}

To investigate the electronic structures of these uniaxial deformed
AGNRs, the deformation of AGNRs is quantified by the strain
($\varepsilon$) defined as $\varepsilon$=(r-r$_{0}$)/r$_{0}$, where
$r$ and $r_{0}$ (r$_{0}$=4.287 {\AA}) is the deformed and initial
equilibrium lattice constant along the axial direction of AGNRs,
respectively.\cite{test} The electronic structures of the deformed
AGNRs with three different widths are all calculated. The band
structures for AGNR with width W=13 under five different uniaxial
strains are shown Figure~\ref{fig2}(a), where
$\varepsilon$=-4.0$\%$, -0.8$\%$, 3.0$\%$, 7.3$\%$, and 10.0$\%$
labeled with A, B, C, D, and E, respectively. Note that B and D
correspond to two turning points (the maximal and minimal energy
gaps), while A, C, E are three intergradation points. Clearly, they
exhibit direct band gaps at $\Gamma$ point for all cases. The
obvious difference among band structures of the deformed AGNRs under
five different given $\varepsilon$ values is the positions of two
upmost valence subbands (v1 and v2) and two lowest conduction
subbands (c1 and c2) relative to the Fermi level. When the applied
strain ($\varepsilon$) is set to be -4.0 $\%$, all subbands (v1, v2,
c1, and c2) are separated at $\Gamma$ point. At the maximal energy
gap point, we observe that two valence subbands (v1 and v2) and both
conduction subbands (c1 and c2) degenerate when the geometric
deformation is about -0.8$\%$. These subbands are separated again
when further elongating the AGNR up to 3.0 $\%$. When the strain
increases continually to 7.3 $\%$ (corresponding to the minimal
energy gap point), the subband v1 shifts upwards while c1 moves down
to the Fermi level, leading these two subbands almost degenerate.
When $\varepsilon$ further increases to 10.0 $\%$, four subbands are
separated again.

We further investigate their electronic properties and observe an
interesting phenomenon. The spatial distributions of these subbands
(v1, v2, c1, or c2) of the deformed AGNRs with given $\varepsilon$
are plotted in Fig. 2 (c)-(d). For A ($\varepsilon$=-4.0 $\%$) case,
the valence subband v1 is mainly contributed by the parallel axial
bonds, while v2 is featured by the vertical axial bonds. At the turning
point B ($\varepsilon$=-0.8 $\%$), it is clear that two subbands
(v1 and v2) degenerate at $\Gamma$ point and their spatial distributions
are swapped as seen in Fig.~\ref{fig2} (b). As increasing the strain
to 3.0 $\%$, the spatial distributions of c1 and c2 are interchanged
as illustrated in Fig.~\ref{fig2} (c). Comparing with spatial pattern
at D ($\varepsilon$=7.3 $\%$), this phenomenon happens to two subbands
c1 and v1 at E ($\varepsilon$=10.0 $\%$) as shown in Fig.~\ref{fig2}(d).

To display more clearly, the variations of energy gap of AGNRs with
width W=12, 13, and 14 as a function of $\varepsilon$ are shown in
Figure~\ref{fig3}(a) with filled square, circle, and triangle symbol
lines, respectively. The calculated maximal values of $E_{g}$ for
the AGNR with widths W=12, 13, and 14 are 1.07, 1.00, and 0.97 eV
appearing at $\varepsilon$=5.0$\%$, -0.8$\%$, and 9.5$\%$,
respectively, while the minimal values of $E_{g}$ are 0.02, 0.03,
and 0.03 eV, which occurs at $\varepsilon$=-4.5$\%$, 7.3$\%$, and
1.3$\%$. Although the exact semiconductor-to-metal transition does
not achieve by both elongating and compressing the AGNRs, it is
clear that the value of energy gap is sensitive to the applied
strain ($\varepsilon$). In other words, the uniaxial strain strongly
affects the electronic structures of AGNRs. This implies that AGNRs
can be used to design as strain sensor. It is interesting to note
that the shapes of calculated curves display zigzag feature for
three different ribbon widths. The energy gaps change almost
linearly between two neighboring turning points by changing the
$\varepsilon$.

The variations of the energy gaps of three family structures with
different widths (W=3n, 3n+1, and 3n+2, where n=4, 5,and 6) as a function
of $\varepsilon$ are shown in Fig.~\ref{fig3} (b), (c), and (d),
respectively. Obviously, there exist following four main common features
for all AGNRs. (1) The zigzag feature is observed for the deformed
AGNRs with large width; (2) The energy gap decreases when the width
of AGNRs increases without geometric deformation; (3) The minimal
energy gap of all deformed AGNRs is several meV, while the maximal
energy gap is sensitive to the width of the deformed AGNRs and its
value reduces as increasing the width of AGNRs. For example, for the
deformed AGNRs with width W=3n (n=4, 6, and 8), the maximal values
of energy gaps are 1.07, 0.74, and 0.56 eV, which appear at $\varepsilon$=9.5
$\%$, 6.6 $\%$, and 4.8 $\%$, respectively, as shown in Fig.~\ref{fig3}
(b). (4) Clearly, the distance between between two turning points
becomes shorter when the width of AGNRs increases, which suggests
that the tunable window of energy gap becomes narrow for the wider
AGNRs.

In general, only including a constant nearest neighbor hopping
integral (t) between $\pi$-electrons TBA results of AGNRs are
different from these by first-principles calculations.\cite{Son}
However, it could reproduce the DFT results of hydrogen passivated
AGNRs through introducing of an additional edge hopping
parameter.\cite{Son,Wang-GNR} Here, to capture a clearer picture,
the electronic structures of the deformed AGNRs are also calculated
using the TB model. According to the geometric optimized results, we
find that there are four kinds of carbon-carbon bond length in the
deformed AGNRs, as labeled by a$_{n}$ (n=1 to 4) in Fig.~\ref{fig1},
where a$_{1}$ and a$_{2}$ stand for the inner C-C distances, while
a$_{3}$ and a$_{4}$ describe the edge C-C separations. The variation
of four kinds of bond lengths as a function of the strain
$\varepsilon$ are shown in Fig.~\ref{fig4}(a) for the deformed AGNRs
with width W=13. Similar results are obtained for AGNRs with width
W=12 and 14. It is clear that the C-C separations change almost
linearly with the strain and the deformation leads to the largest
change of the inner C-C bond length (a$_{1}$). The change of C-C
distance results in variation of the hopping parameter between two
neighbor carbon atoms (t) change in the deformed AGNRs. For
simplicity, we assume that the change of t ($\Delta$t) is
proportional to $\Delta$a linearly. Comparing with the change of
t$_{1}$ ($\Delta$t$_{1}$), the changes of other three hopping
integrals are set to $\Delta$t$_{2}$=0.15$\Delta$t$_{1}$,
$\Delta$t$_{3}$=0.40$\Delta$t$_{1}$, and
$\Delta$t$_{4}$=0.13$\Delta$t$_{1}$, respectively. The coefficient
before $\Delta$t$_{1}$ is determined by the relative slope as shown
in Fig.~\ref{fig4} (a). Four initial hopping parameters for the
equilibrium AGNRs without deformation are set to be: t$_{1}$=-2.7
eV, t$_{2}$=-2.65 eV, t$_{3}$=-3.2 eV, and t$_{4}$=-2.75 eV. The
changes of energy gaps of AGNRs versus t$_{1}$ are shown in
Fig.~\ref{fig4} (b). Clearly, the TBA results reproduce the main
feature of DFT calculations. This result shows that the electronic
structures of AGNRs can be described by introducing the hopping
parameter t$_{1}$ in TB scheme. The change of the hopping parameters
of the deformed AGNRs are responsible for the variation of energy
gaps.

Recently, Han \emph{et al.} have measured the size of the energy gap
of graphene nanoribbons with various widths from 10 to 100
nm.\cite{Melinda} They found that the energy gap scales inversely
with the ribbon width. Here, we extend the TBA calculations to get
the variation of maximal value of $E_{g}$ as a function of the width
of AGNRs (W), as shown in Fig.~\ref{fig4} (c). It is clear that
$E{_{g}}_{max}$ decreases smoothly as increasing the width of
AGNNRs, which is independent of the family structures. By fitting
the calculated curve, we obtain a simple empirical formula,
$E{_{g}}_{max}$ (eV)=14.06/W, as plotted in Fig.~\ref{fig4} (c) with
the green line. We find that this scaling relationship can be used
to calibrate the maximal energy gap of AGNRs with large width. For
example, the maximal value of energy gap of AGNR with W=100 (about
12 nm) is found to be 0.14 eV from TB calculation, remarkably, this
empirical relation curve gives the value of 0.14 eV as well. This
result also agrees well with the reported experimental value of the
GNR without geometric deformation.\cite{Melinda}

In summary, the electronic structures of deformed AGNRs are calculated
by using \textit{ab initio} methods and TB methods. The energy gaps
of AGNRs are predicted to change with zigzag shape as a function of
the applied strain. The tunable window of the energy gap becomes narrower
when the width of ANGRs increases. TBA simulations reproduce these
results by first-principles calculations. We find that the change
of hopping integrals between $\pi$-orbitals of carbon atoms are responsible
for the variation of the energy gap of deformed AGNRs. These findings
are helpful to construct and design graphene nanoelectronic devices
in the near future.

The authors would like to thank Haibin Su for his discussions. This
work was partially supported by the National Natural Science Foundation
of China under Grand Nos. 10674121, 20533030, 10574119, and 50121202,
by National Key Basic Research Program under Grant No. 2006CB922004,
by the USTC-HP HPC project, and by the SCCAS and Shanghai Supercomputer
Center.

\clearpage
\begin{figure}

\caption{(Color online) Schematic of an AGNR with width W=13. Here,
the one dimensional unit cell distance between two dash-dotted lines
is represented by $r$. The blue atoms denote hydrogen atoms
passivate the edge carbon atoms (black dots). Four kinds of
carbon-carbon bond lengths are labeled with a$_{1}$ to a$_{4}$.}

\caption{(Color online) (a) Band structures of AGNRs with the width
W=13. Here, the uniaxial strain ($\varepsilon$) is set to be -4.0,
-0.8, 3.0, 7.3, and 10.0 $\%$ and labeled with A, B, C, D, and E,
respectively. (b), (c) and (c) The spatial distribution of these
subbands (v1, v2, c1, and c2) near Fermi level of the deformed AGNRs
with different strains. }

\caption{(Color online) Variation of the energy gaps (E$_{g}$) for
the AGNRs with width W=12, 13, and 14 as a function of the strain
($\varepsilon$). Variation of $E_{g}$ as a function of $\varepsilon$
for three family structures with different widths, (a) W=3n, (b)
W=3n+1, and (c) W=3n+2 (n=4, 6, and 8.)}

\caption{(Color online) (a) Four kinds of C-C bond lengths of the
deformed AGNRs with width W=13 as a function of $\varepsilon$. (b)
The Variation of the band gap obtained by TBA method for AGNRs with
W=12, 13, an 14 as a function of the nearest neighbor hopping
integral t$_{1}$. (c) The maximal value of energy gap of AGNRs
obtained from TBA results versus the ribbon widths with three family
structures (W=3n, 3n+1, and 3n+2, n is a integer).}

\end{figure}

\clearpage

\begin{figure}
\includegraphics[width=7.5cm]{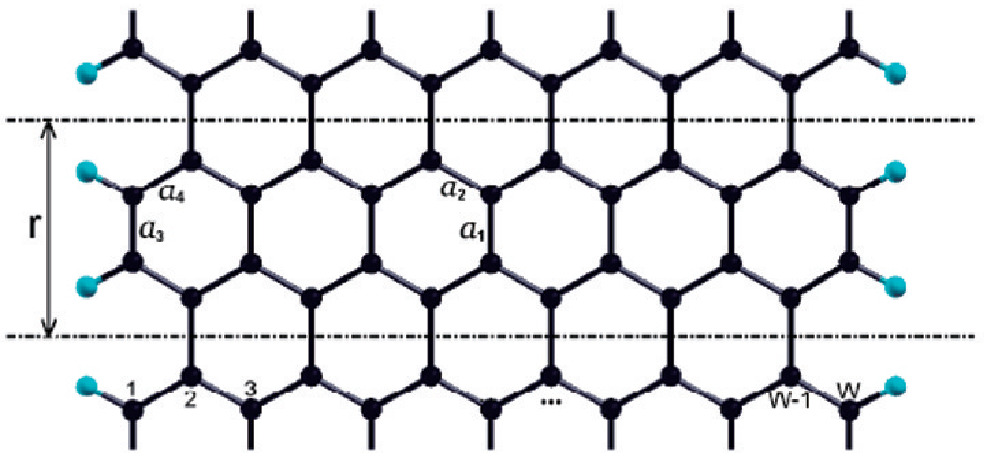}
\hspace{10cm} \centerline{Fig. 1 of Sun \emph{et al.}}
\end{figure}

\clearpage
\begin{figure}
\includegraphics[width=7.5cm]{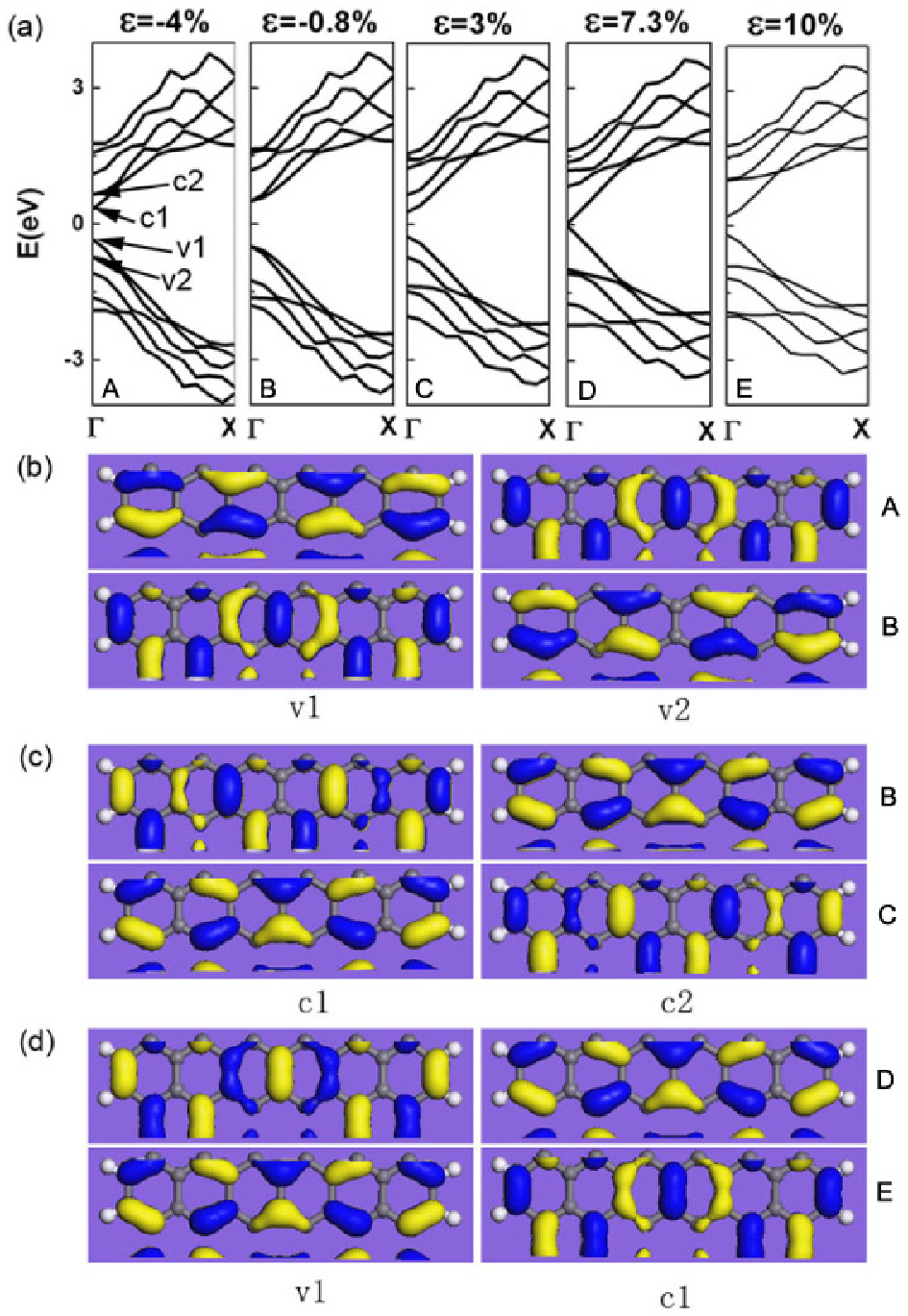}
\hspace{10cm} \centerline{Fig. 2 of Sun \emph{et al.}}
\end{figure}

\clearpage
\begin{figure}
\includegraphics[width=7.5cm]{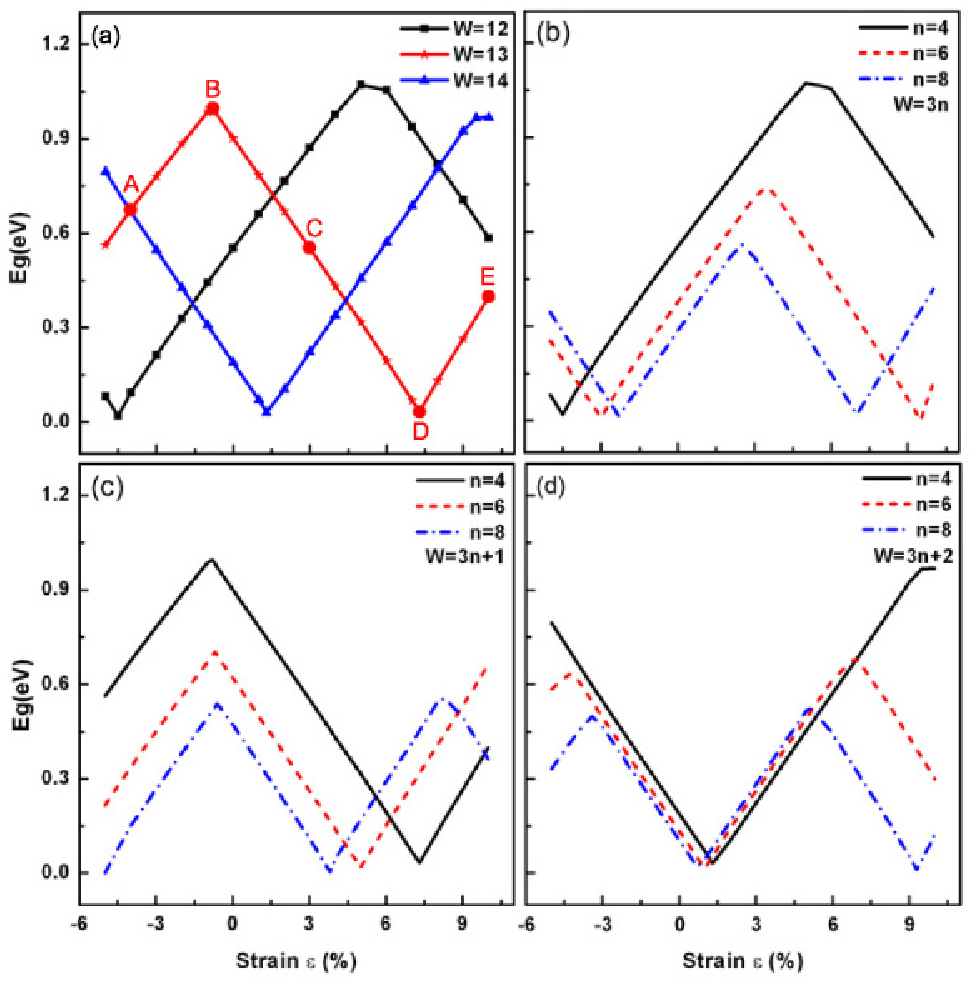}
\hspace{10cm} \centerline{Fig. 3 of Sun \emph{et al.}}
\end{figure}

\clearpage
\begin{figure}
\includegraphics[width=7.5cm]{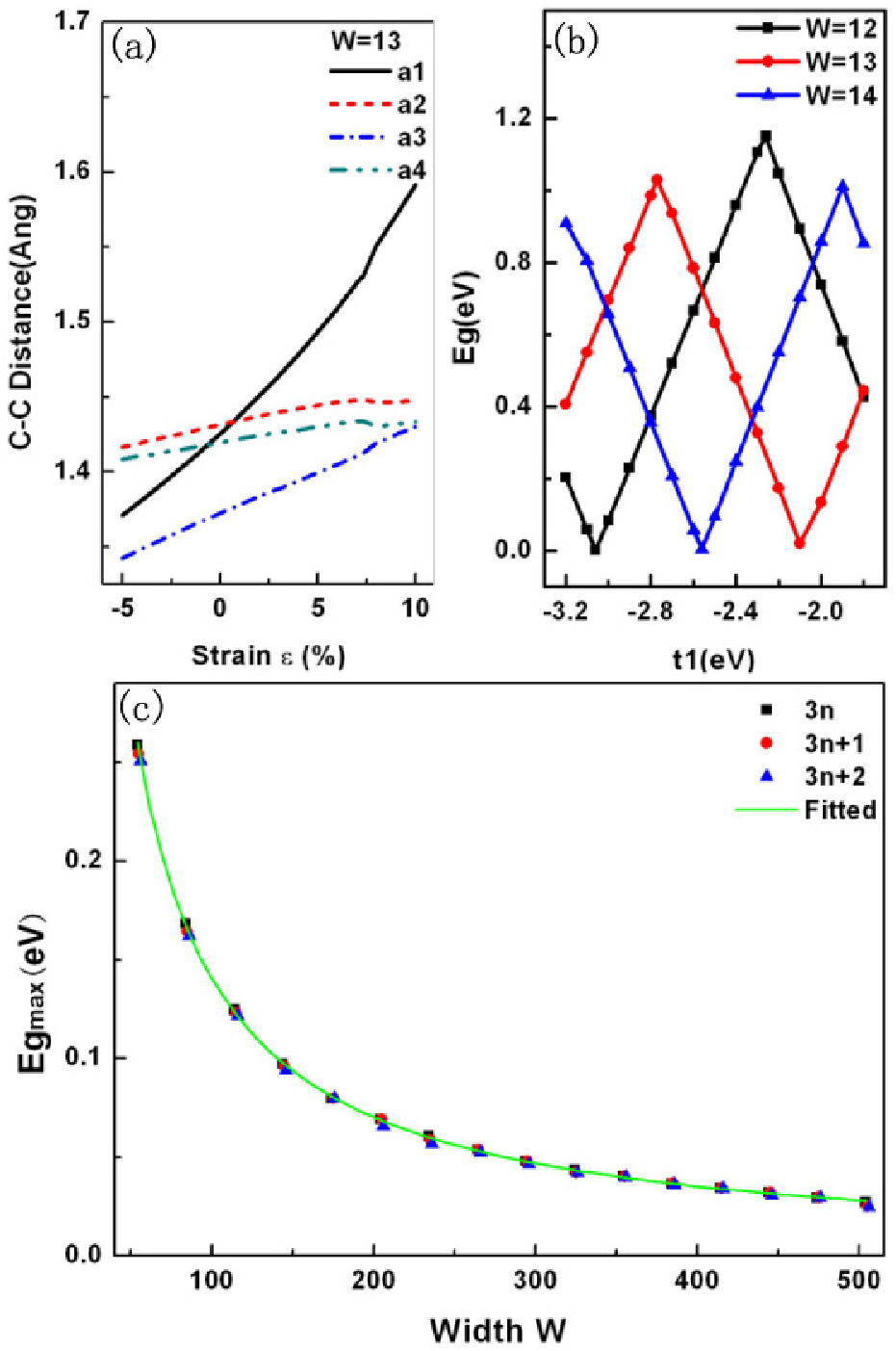}
\hspace{10cm} \centerline{Fig. 4 of Sun \emph{et al.}}
\end{figure}

\end{document}